\documentclass{ws-procs9x6}            

\usepackage{graphicx,color,hyperref}
\usepackage{amsmath}
\usepackage{amsfonts}
\usepackage{amssymb}
\usepackage{braket}
\usepackage{mathtools}
\usepackage{graphicx}
\usepackage{float}
\usepackage{subfigure}

\begin{document}
\title{Three Dimensional Imaging of Proton in Basis Light-Front Quantization}

\author{Siqi Xu$^{1,2*}$, Chandan Mondal$^{1,2+}$, Jiangshan Lan$^{1,2}$, Xingbo Zhao$^{1,2**}$, Yang Li$^{3}$, James P. Vary$^{3}$\\(BLFQ Collaboration)}

\address{$1$.Institute of Modern Physics, Chinese Academy of Science,\\ Lanzhou 730000, China\\
$2$.School of Nuclear Science and Technology, University of Chinese Academy of Sciences, Beijing 100049, China\\
$3$.Department of Physics and Astronomy, Iowa State University,\\ Ames, Iowa 50011, USA\\
$^*$E-mail: xsq234@impcas.ac.cn\\
$^+$E-mail: mondal@impcas.ac.cn\\
$^{**}$E-mail: xbzhao@impcas.ac.cn}





\begin{abstract}
We employ an effective Hamiltonian that includes the transverse and longitudinal confinement and the one-gluon exchange interaction with fixed coupling constant.  By solving the eigenvalue equation in basis light-front quantization (BLFQ), we generate the light-front wavefunctions (LFWFs) for the nucleon in the valence quark Fock space. Fitting the model parameters, we obtain high quality descriptions of electromagnetic form factors and radius for proton while the results deviate somewhat from experimental data for neutron.
\end{abstract}

\keywords{BLFQ, LFWF, FFs, PDFs, GPDs}

\bodymatter

\section{Introduction}\label{aba:sec1}

Basis light-front quantization (BLFQ) is a nonperturbative approach which is developed for solving bound state problems in quantum field theories\cite{Vary:2009gt,Wiecki:2014ola,Honkanen:2010rc,Li:2017mlw,Li:2015zda,Chakrabarti:2014cwa,Lan:2019vui,Tang:2018myz,Xu:2019xhk,Du:2019ips,Adhikari:2016idg,Adhikari:2018umb,Li:2019kpr}. This approach has been successfully applied to QED\cite{Wiecki:2014ola,Chakrabarti:2014cwa} and QCD\cite{Li:2017mlw,Li:2015zda,Lan:2019vui,Tang:2018myz,Xu:2019xhk,Du:2019ips,Adhikari:2016idg,Adhikari:2018umb,Li:2019kpr} systems. In our work, we apply the BLFQ approach to the nucleon and study the electromagnetic form factors. As a Hamiltonian formalism, we adopt a light-front effective Hamiltonian, which includes the holographic QCD confinement potential supplemented by longitudinal confinement\cite{Li:2017mlw,Li:2015zda,Brodsky:2014yha} along with the one-gluon exchange interaction with a fixed coupling constant. The light-front wave functions (LFWFs) are obtained by diagonalizing the effective Hamiltonian and used to calculate the electromagnetic form factors.

Electromagnetic form factors are crucial for probing the structure of the nucleon. In the light-front formalism, the Dirac and Pauli form factors, $F_1(Q^2)$ and $F_2(Q^2)$, are defined with the longitudinal vector current ($J^+$)\cite{Brodsky:1980zm,Cates:2011pz}
\begin{eqnarray}
\braket{P+q,\uparrow|\frac{J^+(0)}{2P^+}|P,\uparrow} &=& F_1(Q^2), \\
\braket{P+q,\uparrow|\frac{J^+(0)}{2P^+}|P,\downarrow} &=& -(q^1-iq^2)\frac{F_2(Q^2)}{2M},
\end{eqnarray}
where $Q^2=-q^2=q_{\perp}^2$ is the square of the momentum transfer, and $M$ is the nucleon mass. The ket $\ket{P,S_z}$ represents the physical state that can be expanded in terms of the wave functions\cite{Brodsky:2014yha},
\begin{eqnarray}
\ket{P,S_z} = &&\!\!\!\! \int \prod_{i=1}^{3} \frac{dx_id^2k_{i\perp}}{\sqrt{x_i}16\pi^3} 16\pi^3\delta \left(1-\sum_{i=1}^{3} x_i\right) \delta^2 \left(\sum_{i=1}^{3}k_{i\perp}\right) \nonumber \\
&&\!\!\!\! \times \Psi^{\Lambda}(x_i,k_{i\perp},\lambda_i) \ket{x_iP^+_i,k_{i\perp}+x_iP_{\perp},\lambda_i}.\label{wavefunction_expansion}
\end{eqnarray}
Here, the $S_z$ and $\lambda_i$ are helicities of the nucleon and quarks respectively. The $x_i=\frac{k_i^+}{P^+}$ is the longitudinal momentum fraction of quarks.
Thus, the flavor form factors can be written as the overlap of light-front wave functions.
The nucleon Sachs form factors are written in the terms of Dirac and Pauli form factors,
\begin{eqnarray}
G_E^i(Q^2)= F_1^{i}(Q^2) - \frac{Q^2}{4*M_i^2} F_2^{i}(Q^2), ~~~~
G_M^i(Q^2)= F_1^{i}(Q^2) + F_2^i(Q^2).
\end{eqnarray}
The $i = \rm{P ~or~ N }$ represents the proton or neutron, and $F_{1/2}^i=\sum_f e_f F_{1/2}^{f/i}$ is the Dirac (Pauli) form factors of the nucleon~\cite{Beringer:1900zz}. And the electromagnetic radii of the nucleon can be obtained from
\begin{eqnarray}
\braket{r^2_E}^i=-6 \frac{dG^i_E(Q^2)}{dQ^2}\bigg|_{Q^2=0}, ~~~~
\braket{r^2_M}^i=-\frac{6}{G^i_M(0)}\frac{dG^i_M(Q^2)}{dQ^2}\bigg|_{Q^2=0}.
\end{eqnarray}

\section{Hamiltonian Formalism}

BLFQ solves the eigenvalue equation of the light-front Hamiltonian
$P^- \ket{\beta}= P^-_{\beta} \ket{\beta}$, 
which leads to the eigenvalue $P^-_{\beta}$ and the associated eigenvectors of the bound state.
In our work, we only consider the lowest Fock-sector for the expansion of the nucleon, and employ an effective Hamiltonian $P^-_{\rm{eff}}$ which is given by
\begin{eqnarray}
P^-_{\rm{eff}} =&&\sum_{i} \frac{\rm{m}_i^2+p_{i\perp}^2}{x_i}+\frac{1}{2}\sum_{i,j} \big(\kappa_T^4 x_{i} x_{j}r_{ij\perp}^2+\frac{\kappa_L^4}{(\rm{m}_i+\rm{m}_j)^2}\partial_{x_i}(x_ix_j\partial_{x_j}) \big) \nonumber\\
&& + \frac{1}{2}\sum_{i,j} \frac{C_F4\pi \alpha_s}{Q^2} \bar{u}_{s^{\prime}_i}(k^{\prime}_i)\gamma^{\mu}u_{s_i}(k_i)\bar{u}_{s^{\prime}_j}(k^{\prime}_j)\gamma_{\mu}u_{s_j}(k_j),
\end{eqnarray}
where the $\rm{m}_{i/j}$ is the constituent mass of quarks and the $i,j=1,2,3$ label the Fock particles.
For Each single-particle basis state, we employ the discrete plane-wave basis ($k$) in the longitudinal direction and 2D harmonic oscillator (2DHO) basis ($n$ and $m$) in the transverse direction. Besides, a single quantum number ($\lambda$) presents the helicity degree of freedom.

For the nucleon, proton (or neutron) is the lowest eigenstate, denoted by $\ket{P^{\Lambda}}$, where the $\Lambda$ indicates helicity of the nucleon. In momentum space, the LFWFs are written as
\begin{eqnarray}
\Psi^{\Lambda}&&\!\!\!(x_i,k_{i\perp},\lambda_i)=\sum_{\substack{n_1,m_1,n_2 \\ m_2,n_3,m_3}} \big( \psi^{\Lambda}(k_{i},n_{i},m_{i},\lambda_i) \nonumber \\&&
\times \prod_i \frac{\sqrt{2}}{b(2\pi)^{\frac{3}{2}}}\sqrt{\frac{n!}{(n+|m|)!}}e^{-p_{\perp}^2/(2b^2)}
  \left(\frac{|p_{\perp}|}{b}\right)^{|m|}L^{|m|}_{n}(\frac{p_{\perp}^2}{b^2})e^{im\theta}\big).
\end{eqnarray}
Here, b is an HO basis parameter with the dimension of mass, and $L^{|m|}_{n}(\frac{p_{\perp}^2}{b^2})$ is the generalized Laguerre polynomial.

\section{Numerical Results}

In this paper, we set the model parameters $m_{q/\rm{OGE}}=0.2~\rm{GeV}$, $m_{q/k}=0.3~\rm{GeV}$, ~$\kappa_T=0.284~\rm{GeV}$, $\kappa_L=0.373~\rm{GeV}$ and $\alpha_s=1.0 \sim 1.2$.
\begin{figure*}[htbp]
\centering
\subfigure[]{
\begin{minipage}[t]{0.45\linewidth}
\centering
\includegraphics[width=\columnwidth]{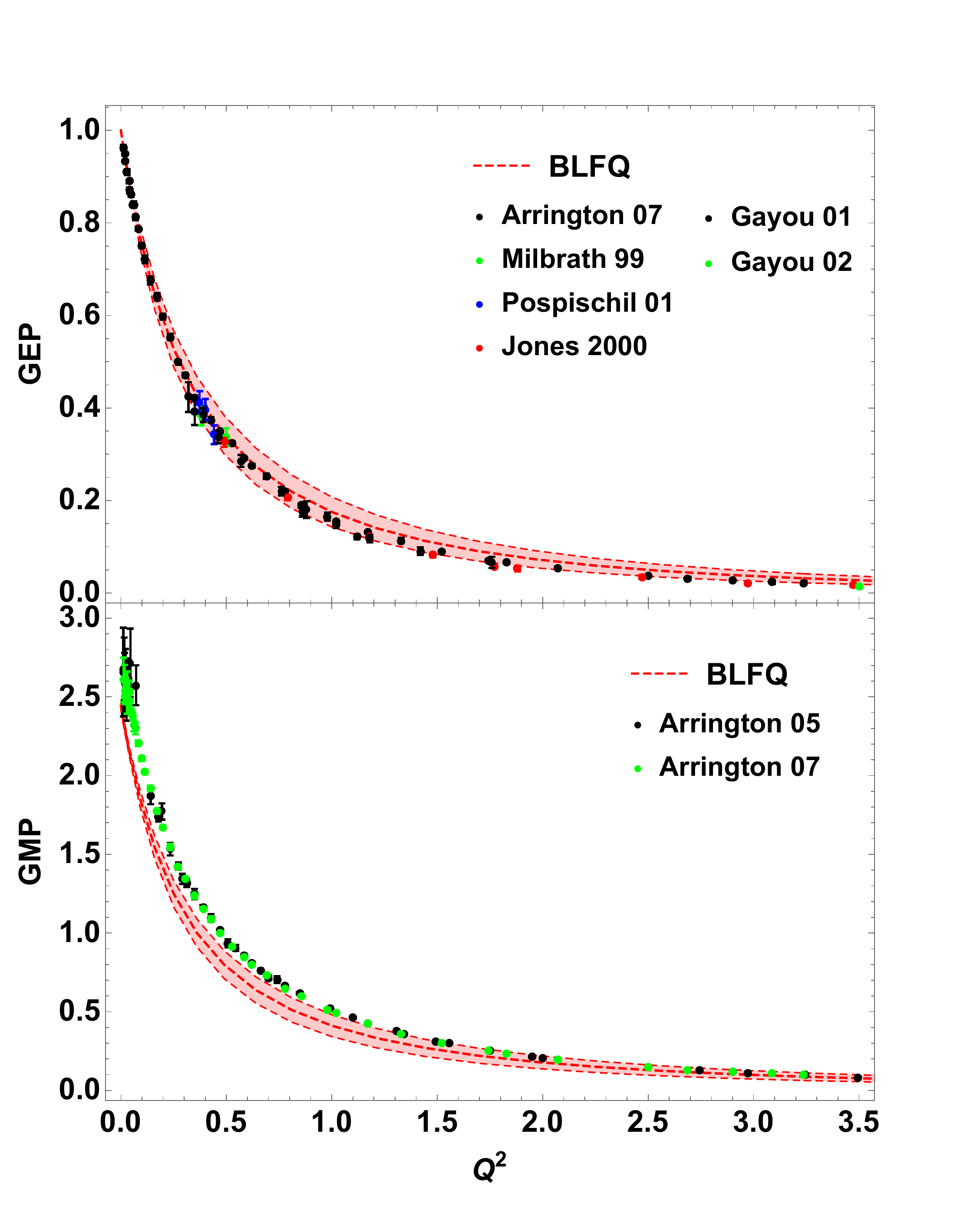}
\end{minipage}
\label{sach_proton}
}
\subfigure[]{
\begin{minipage}[t]{0.45\linewidth}
\centering
\includegraphics[width=\columnwidth]{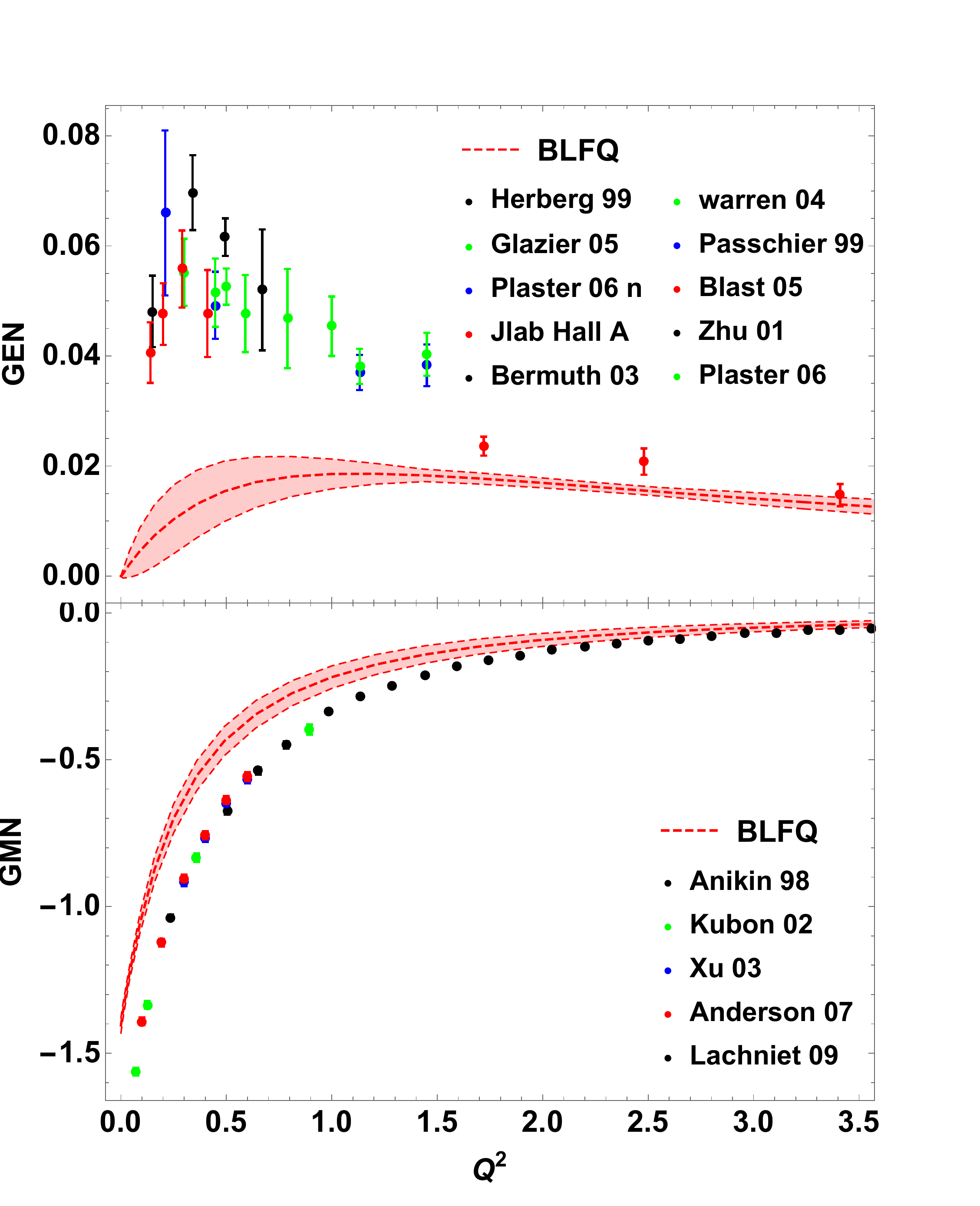}
\end{minipage}
\label{sach_neutron}
}
\caption{The Sachs form factors for the proton (a) and neutron (b). Nucleon Sach's FFs $G^{P/N}_E(Q^2)$ (upper panel) and $G^{P/N}_M(Q^2)$ (lower panel) are functions of $Q^2$. The bands are BLFQ results reflecting our $\alpha_s$ uncertainty of $10\%$. The experimental data are taken from Ref~\cite{Chakrabarti:2013dda}.}
\label{sach}
\end{figure*}
In Fig~\ref{sach_proton}, the Sachs form factors of the proton show an agreement with the experimental data, Except for $G^{\rm{P}}_M$ in the low $Q^2$ region.  At $Q^2=0$, $G_M(0)$ gives the anomalous magnetic moments. Our calculations show that the anomalous magnetic moments of the proton ($G^{\rm{P}}_M(0)=2.443\pm0.027$) is somewhat different with the experimental measurements ($G^{\rm{P}}_M(0)=2.79$). In Fig~\ref{sach_neutron}, we show the Sachs form factors of the neutron and compare them with the experimental data revealing a significant difference. Especially, at $Q^2= 0$, the $G_M^{\rm{N}}(0)= -1.405\pm0.026$ is disagrees with the experimental data ($G_M^{\rm{N}}(0)=-1.91$). 

\begin{table}
\tbl{Electromagnetic radii of the nucleon. Our results are compared with the experimental data~\cite{Beringer:1900zz}.}
{\begin{tabular}{@{}ccccc@{}}\toprule
      ~&~$\braket{r_E^P}/(\rm{fm})$~&~$\braket{r_M^P}/(\rm{fm})$~&~$\braket{r_E^N}^2/(\rm{fm}^2)$~&~$\braket{r_M^N}/(\rm{fm})$    \\
\colrule
 BLFQ     ~&~  $0.85\pm0.05$    ~&~ $0.88\pm0.03$     ~&~$-0.09\pm0.17$     ~&~$0.90\pm0.03$      \\
 Exp. Data~&~  $0.833\pm 0.010$   ~&~ $0.777\pm 0.016$  ~&~$-0.1161\pm 0.0022$    ~&~$0.862^{+0.009}_{-0.008}$\\
\botrule
\end{tabular}
}
\label{tab:radii}
\end{table}

We also calculate the electromagnetic radii of the nucleons, which we show in Table~\ref{tab:radii}. The BLFQ results are in a good agreement with the experimental data\cite{Beringer:1900zz}.

\section{Conclusion}

In our work, we produce the light-front wave functions by solving the eigenvalue equation of light-front Hamiltonian, and evaluate the electromagnetic form factors of the nucleon. We observe the proton form factors are in a reasonable agreement with the experimental data. The neutron form factors show a significant issue in the low Q region. We also compare the electromagnetic radii of the nucleon with the experimental data.

\section{Acknowledgment}
We thank Henry Lamm, Wei Zhu for many useful discussions. CM is supported by the National Natural Science Foundation of China (NSFC) under the Grant No. 11850410436. XZ is supported by new faculty startup funding by the Institute of Modern Physics, Chinese Academy of Sciences and by Key Research Program of Frontier Sciences, CAS, Grant No ZDBS-LY-7020. JPV is supported by the Department of Energy under Grants No. DE-FG02-87ER40371, and No. DE-SC0018223 (SciDAC4/NUCLEI). A portion of the computational resources were provided by the National Energy Research Scientific Computing Center (NERSC), which is supported by the Office of Science of the U.S. Department of Energy under Contract No.DE-AC02-05CH11231.


\begin{thebibliography}{99}
\bibitem{Vary:2009gt} 
  J.~P.~Vary {\it et al.},
  Phys.\ Rev.\ C {\bf 81}, 035205 (2010)
\bibitem{Honkanen:2010rc} 
  H.~Honkanen {\it et al.}, 
Phys.\ Rev.\ Lett.\  {\bf 106}, 061603 (2011)
\bibitem{Wiecki:2014ola} 
  P.~Wiecki, Y.~Li, X.~Zhao, P.~Maris and J.~P.~Vary,
  Phys.\ Rev.\ D {\bf 91}, no. 10, 105009 (2015)
\bibitem{Chakrabarti:2014cwa} 
  D.~Chakrabarti, X.~Zhao, H.~Honkanen, R.~Manohar, P.~Maris and J.~P.~Vary,
  Phys.\ Rev.\ D {\bf 89}, no. 11, 116004 (2014)
\bibitem{Li:2017mlw} 
  Y.~Li, P.~Maris and J.~P.~Vary,
  Phys.\ Rev.\ D {\bf 96}, no. 1, 016022 (2017)
\bibitem{Li:2015zda} 
  Y.~Li, P.~Maris, X.~Zhao and J.~P.~Vary,
  Phys.\ Lett.\ B {\bf 758}, 118 (2016)
\bibitem{Lan:2019vui} 
  J.~Lan, C.~Mondal, S.~Jia, X.~Zhao and J.~P.~Vary,
  Phys.\ Rev.\ Lett.\  {\bf 122}, no. 17, 172001 (2019)
\bibitem{Tang:2018myz} 
  S.~Tang, Y.~Li, P.~Maris and J.~P.~Vary,
  Phys.\ Rev.\ D {\bf 98}, no. 11, 114038 (2018)
\bibitem{Xu:2019xhk} 
  C.~Mondal, S.~Xu, J.~Lan, X.~Zhao, Y.~Li, D.~Chakrabarti and J.~P.~Vary,
  arXiv:1911.10913 [hep-ph].
\bibitem{Du:2019ips} 
  W.~Du, Y.~Li, X.~Zhao, G.~A.~Miller and J.~P.~Vary,
  arXiv:1911.10762 [nucl-th].
\bibitem{Adhikari:2016idg} 
  L.~Adhikari, Y.~Li, X.~Zhao, P.~Maris, J.~P.~Vary and A.~Abd El-Hady,
  Phys.\ Rev.\ C {\bf 93}, no. 5, 055202 (2016)
\bibitem{Adhikari:2018umb} 
  L.~Adhikari, Y.~Li, M.~Li and J.~P.~Vary,
  Phys.\ Rev.\ C {\bf 99}, no. 3, 035208 (2019)
\bibitem{Li:2019kpr} 
  M.~Li, Y.~Li, P.~Maris and J.~P.~Vary,
  Phys.\ Rev.\ D {\bf 100}, no. 3, 036006 (2019)
\bibitem{Brodsky:2014yha} 
  S.~J.~Brodsky, G.~F.~de Teramond, H.~G.~Dosch and J.~Erlich,
  Phys.\ Rept.\  {\bf 584}, 1 (2015)
\bibitem{Brodsky:1980zm} 
  S.~J.~Brodsky and S.~D.~Drell,
  Phys.\ Rev.\ D {\bf 22}, 2236 (1980).
\bibitem{Cates:2011pz} 
  G.~D.~Cates, C.~W.~de Jager, S.~Riordan and B.~Wojtsekhowski,
  Phys.\ Rev.\ Lett.\  {\bf 106}, 252003 (2011)
\bibitem{Beringer:1900zz} 
  J.~Beringer {\it et al.} [Particle Data Group],
  Phys.\ Rev.\ D {\bf 86}, 010001 (2012).
\bibitem{Chakrabarti:2013dda} 
  D.~Chakrabarti and C.~Mondal,
  Eur.\ Phys.\ J.\ C {\bf 73}, 2671 (2013)


\end{thebibliography}
\end{document}